\pdfoutput=1
\documentclass[
reprint,
groupedaddress,
showpacs,
bibnotes,
amsmath,amssymb,
prb,
]{revtex4-1}

\usepackage{graphicx}% Include figure files
\usepackage{dcolumn}% Align table columns on decimal point
\usepackage{bm}% bold math
\usepackage{setspace}

\begin{document}
\newcommand{\BFA}{BaFe$_{2}$As$_{2}$}
\newcommand{\BKFA}{(Ba$_{1-x}$K$_x$)Fe$_{2}$As$_{2}$}
\newcommand{\CFPPA}{Ca$_{10}$(Fe$_{1-x}$Pt$_x$As)$_{10}($Pt$_{3+y}$As$_8$)}
\newcommand{\CFPA}{Ca$_{10}$(FeAs)$_{10}($Pt$_3$As$_8$)}
\newcommand{\CFPvA}{Ca$_{10}$(FeAs)$_{10}($Pt$_4$As$_8$)}

%\preprint{APS/123-QED}

\title{The role of different negatively charged layers in {\CFPPA} and superconductivity at 30 K in electron-doped (Ca$_{0.8}$La$_{0.2}$)$_{10}$(FeAs)$_{10}$(Pt$_3$As$_8$)}

\author{Tobias St\"{u}rzer}
\author{Gerald Derondeau}
\author{Dirk Johrendt}%
\email{johrendt@lmu.de}\affiliation{Department Chemie, Ludwig-Maximilians-Universit\"{a}t M\"{u}nchen, Butenandtstra\ss e 5-13 (Haus D), 81377 M\"{u}nchen, Germany\\}

\date{\today}%

\begin{abstract}

The recently discovered compounds {\CFPPA} exhibit superconductivity up to 38 K, and contain iron arsenide (FeAs) and platinum arsenide (Pt$_{3+y}$As$_8$) layers separated by layers of Ca atoms. We show that high $T_c$'s above 15~K only emerge if the iron-arsenide layers are at most free of platinum-substitution ($x \rightarrow$ 0) in contrast to recent reports. In fact Pt-substitution is detrimental to higher $T_c$, which  increases up to 38~K only by charge doping of pure FeAs layers.
We point out, that two different negatively charged layers [(FeAs)$_{10}$]$^{n-}$ and (Pt$_{3+y}$As$_8$)$^{m-}$ compete for the electrons provided by the Ca$^{2+}$-ions, which is unique in the field of iron-based superconductors. In the parent compound {\CFPA}, no excess charge dopes the FeAs-layer, and superconductivity has to be induced by Pt-substitution, albeit below 15~K. In contrast, the additional Pt-atom in the Pt$_4$As$_8$ layer shifts the charge balance between the layers equivalent to charge doping by 0.2 electrons per FeAs. Only in this case $T_c$ raises to 38~K, but decreases again if additionally platinum is substituted for iron. This charge doping scenario is supported by our discovery of superconductivity at 30~K in the electron-doped La-1038 compound (Ca$_{0.8}$La$_{0.2}$)$_{10}$(FeAs)$_{10}$(Pt$_3$As$_8$) without significant Pt-substitution.

\end{abstract}


\maketitle

%Introduction

\begin{figure}[h]
\center{
\includegraphics[width=60mm]{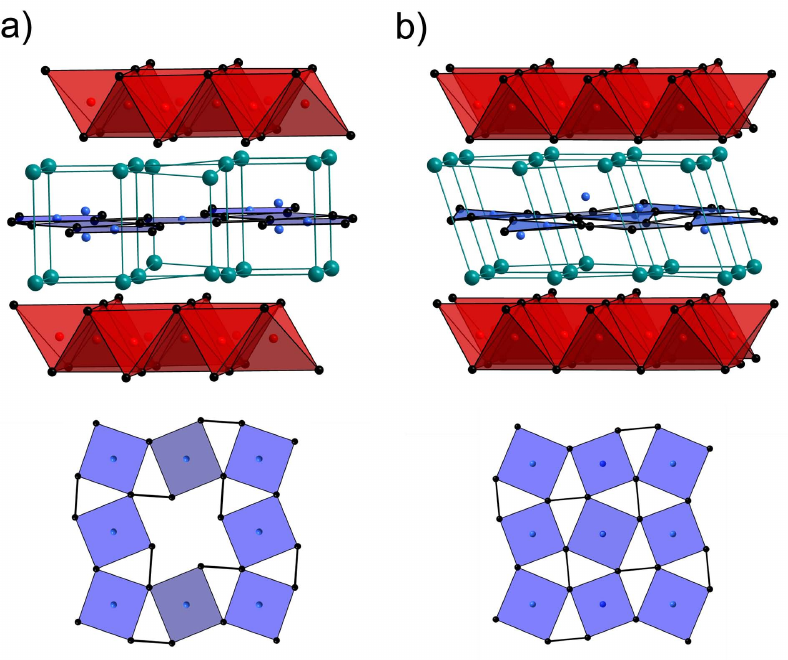}
\caption{Crystal structures of (a) {\CFPA} (1038) and (b) {\CFPvA} (1048).}
\label{fig:structures}
}
\end{figure}

The chemical complexity of the iron-arsenide superconductors has been increased by the recent discovery of the compounds {\CFPPA}.\cite{Loehnert-2011,Cava-PNAS2011,Nohara-2011} Their crystal structures contain alternating layers of iron-arsenide and platinum-arsenide, each separated by calcium atoms (\ref{fig:structures}). Platinum in the Pt$_{3+y}$As$_8$-layers is nearly planar fourfold coordinated by arsenic that forms As$^{4-}_2$-Zintl ions. Two branches of the structural motive have been found depending on the composition of the platinum-arsenide layers. The compound referred to as the 1038-phase contains Pt$_3$As$_8$-layers (Fig.~\ref{fig:structures}a), while in the 1048-phase one more platinum atom is located in Pt$_4$As$_8$-layers (Fig.~\ref{fig:structures}b). The 1038-compound is triclinic, while we have identified three polymorphs of the 1048-phase with tetragonal ($\alpha$-1048, $P4/n$), triclinic ($\beta$-1048, $P\overline{1}$) or monoclinic ($\gamma$-1048, $P2_1/n$) space group symmetries by single crystal X-ray diffraction.

High critical temperatures ($T_c$) up to 38~K have been assigned to the 1048-variants, while $T_c$ of the 1038-phase is below 15~K so far. Recent reports suggested that the critical temperatures are solely controlled by substitution of platinum for iron in the FeAs-layers as known from  Sr(Fe$_{1-x}$Pt$_x$)$_2$As$_2$.\cite{Nishikubo-2010} Nohara~\textit{et al.}\cite{Nohara-2012} even proposed that 'heavy Pt-doping' is required to achieve high $T_c$ in Ca$_{10}$(Fe$_{1-x}$Pt$_x$As)$_{10}$(Pt$_4$As$_8$). Also  Ni~\textit{et al.}\cite{Cava-PNAS2011} have suggested that platinum-substitution controls $T_c$, but the higher values of the 1048-phases were associated with stronger interlayer coupling by Pt--As bonds between the Pt$_4$As$_8$- and FeAs-layers.

However, we have proposed another scenario.\cite{Loehnert-2011} Our band structure calculations  indicated that the Pt$_{3+y}$As$_8$-layers hardly contribute at the Fermi-energy, which is supported by recent angle resolved photoemission experiments, showing that the Fermi-surface topology is similar to those of known FeAs-materials.\cite{Neupane-2011} Thereby it is  extremely unlikely, that critical temperatures as high as 38~K occur in Pt-doped materials, while all hitherto known transition-metal doped FeAs-superconductors remain well below 25~K.

In this letter we show that high critical temperatures in the platinum-iron arsenides are not achieved by Pt-substitution inside the iron-layers, but by charge doping of FeAs-layers. The $T_c(x)$ phase diagrams the 1038 and 1048 compounds are quite different and reveal that Pt-substitution induces superconductivity at low temperatures in the 1038 materials, but is detrimental to $T_c$ in the 1048 compounds, where the FeAs layers are doped by electrons due to a shift of the the charge balance between [(FeAs)$_{10}]^{n-}$ and (Pt$_{3+y}$As$_8$)$^{m-}$. This interpretation is supported by the observation of superconductivity at 30~K in the electron-doped 1038 compound (Ca$_{1-x}$La$_x$)$_{10}$(FeAs)$_{10}($Pt$_3$As$_8$).

%\section{Experimental}

Polycrystalline samples of the platinum-iron arsenides
%{\CFPPA} and [(Ca$_{1-x}$La$_x$)FeAs]$_{10}$Pt$_3$As$_8$ ($x \approx$ 0.2)
were synthesized by solid-state methods from the elements as described in Ref.~\cite{Loehnert-2011}, and characterized by X-ray powder diffraction (PXRD) using the Rietveld method with TOPAS.\cite{Topas} Compositions were determined within errors of $\pm$10\,\%  by refining the occupation parameters and by energy dispersive X-ray spectroscopy (EDX). AC-susceptibility measurements (3~Oe, 1333~Hz) were used to detect superconductivity and the critical temperatures. Full-potential DFT-calculations using the WIEN2k-package\cite{Blaha-2001,Schwarz-2003} together with the QTAIM-method\cite{Bader-1990} were used to calculate and analyze the electron density distribution of the tetragonal 1048-compound.\\

\begin{figure}[h]
\center{
\includegraphics[width=80mm]{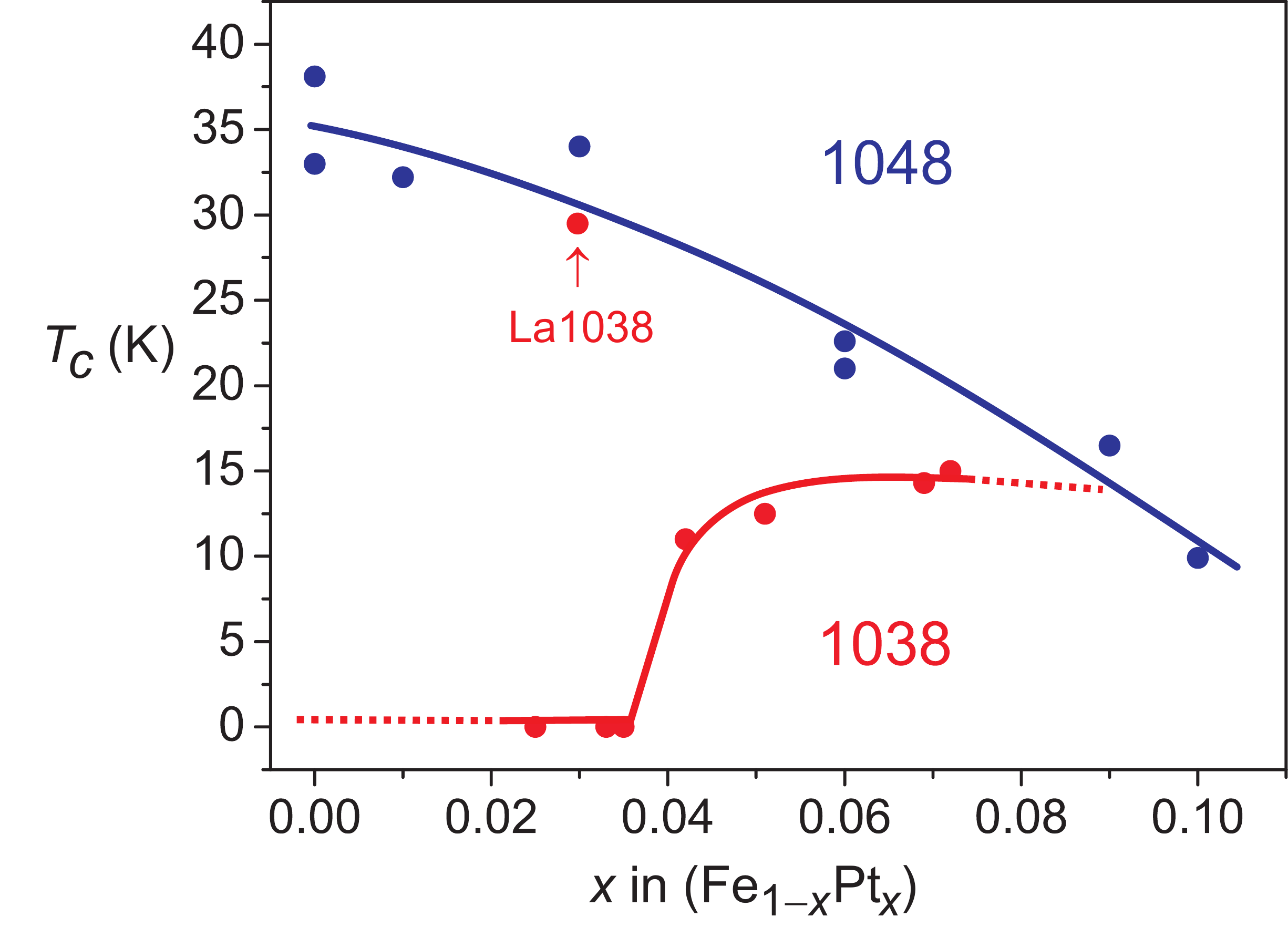}
\caption{Critical temperatures of samples with 1048-structure (blue) and 1038-structure (red). Lines are guides to the eye.}.
\label{fig:phasedia}
}
\end{figure}

Figure~\ref{fig:phasedia} shows the critical temperatures of all samples plotted against the amount of Pt-substitution the iron site ($x$). Compounds with the 1048-structure are well separated from those with 1038-structure. The 1038-compounds  are not superconducting below $x \approx$ 0.03, then $T_c$ increases rapidly up to 15~K. This is very similar to phase diagrams of other FeAs-materials, where superconductivity emerges after suppression of a spin-density-wave (SDW) state of a parent compound by substitution or pressure. Our phase diagram of the 1038-compounds agrees with that in a recent report,\cite{Cho-2011} where structural transitions have been suggested at $x < $ 0.025. Even though the crystal structure of the low temperature phases are still unknown, we refer to  {\CFPA} as the parent compound. Thus far, the 1038-compounds are in line with other FeAs-compounds that become superconducting when doped with transition metals at the iron site, albeit at low $T_c$.

In contrast to this, the critical temperatures of the 1048-compounds are the highest without platinum at the iron site ($x$ = 0), and decrease with the substitution level as shown in Figure~\ref{fig:phasedia}. This is in accordance with the fact, that critical temperatures above 25~K occur in compounds with \textit{charge-doped} FeAs-layers like $Ln$FeAs(O$_{1-x}$F$_x$)\cite{Ren-2008} or {\BKFA},\cite{Rotter-2008} but not in materials with transition-metal substitution at the iron site like Ba(Fe$_{1-x}$Co$_x$)$_2$As$_2$.\cite{Sefat-2008} Thus, charge doping of the FeAs-layer would explain the higher $T_c$ of the 1048-compound, and also the decrease of $T_c$ due to overdoping by additional Pt-substitution at the iron site (Figure~\ref{fig:phasedia}).

It is important to point out, that the platinum-iron-arsenides are the first compounds in this family with \textit{two different negatively charged layers}, and the question arises, whether the electrons provided by the ten Ca$^{2+}$ ions are equally distributed to the [(FeAs)$_{10}$]$^{n-}$ and (Pt$_4$As$_8$)$^{m-}$ layers, respectively. This charge balance is obviously present in the 1038-compound that shows the typical properties of (FeAs)$^{1-}$ layers in other compounds. From this we infer that the 1048-compound experiences a charge imbalance between FeAs and Pt$_4$As$_8$, \textit{viz.} charge doping of the iron arsenide layer. Thus the very different behavior of the critical temperatures is caused by distinct doping scenarios according to the following scheme:

\begin{center}

Ca$^{2+}_{10}$[(FeAs)$_{10}$]$^{10-}$(Pt$_3$As$_8$)$^{10-}$\\
~\\

{\small{Pt-substitution}}~~$\swarrow$~~~~~~~~~~~~~~~~~~$\searrow$~~{\small{charge doping}}\\
~\\

${\rm {Ca}_{10}}({\rm Fe}_{1-x}{{\rm Pt}_x}{\rm {As}})_{10}({\rm {Pt}}_3{\rm {As}}_8)$~
${\rm {Ca}}^{2+}_{10}\rm{[(FeAs)}_{10}]^{12-}({\rm {Pt}}_{4}{\rm {As}}_8)^{8-}$\\
~\\

\end{center}

We suggest that the charge of the FeAs layer increases by the electrons formally provided by the additional Pt$^{2+}$ in the Pt$_4$As$_8$ layer, which means electron doping by 0.2 electrons per FeAs. This is remarkably similar to the electron-doped in the 1111-compounds $Ln$FeAs(O$_{1-x}$F$_x$), which exhibit the highest $T_c$ at $x$ = 0.15-0.2. Our reasoning is supported by the QTAIM-analysis of the charge density distribution in the teragonal 1048-compound. Summing up the charges of the layer atoms results in (Ca$_{10}$)$^{13.6+}$[(FeAs)$_{10}$]$^{7.8-}$(Pt$_4$As$_8$)$^{5.8-}$, thus the electrons are far from being equally distributed. Note also that normalization to Ca$^{2+}$ gives (Ca$_{10}$)$^{20+}$[(FeAs)$_{10}$]$^{11.5-}$(Pt$_4$As$_8$)$^{8.5-}$, which is near to the above mentioned doping scheme. Nevertheless we want to remark that the term atomic charge has no concrete definition in quantum chemistry. For this reason we cannot consider the above values as quantitative.

\begin{figure}
\center{
\includegraphics[width=80mm]{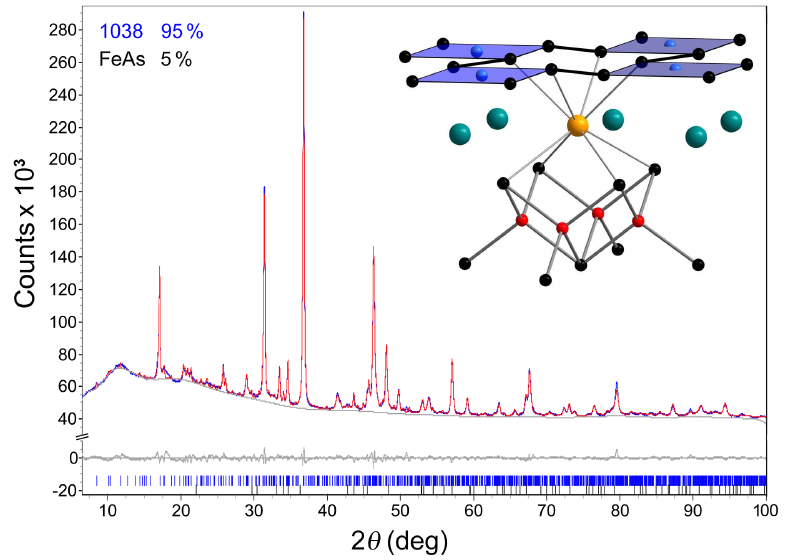}
\caption{X-ray diffraction pattern and Rietveld-fit of (Ca$_{0.8}$La$_{0.2}$)(Fe$_{1-x}$Pt$_x$)$_{10}$(Pt$_3$As$_8$) (La-1038; $P\overline{1}$, $a$ = 8.7493(3) {\AA}, $b$ = 8.7533(2) {\AA}, $c$ = 10.7139(3) {\AA}, $\alpha$ = 75.877(3)$^{\circ}$, $\beta$ = 85.295(3)$^{\circ}$, $\gamma$ = 90.031(3)$^{\circ}$, $R_{wp}$ = 0.016)  Insert: Detail of the crystal structure showing the eightfold coordination of the lanthanum atom by arsenic atoms from the FeAs and Pt$_3$As$_8$ layers, respectively.}
\label{fig:rietveld}
}
\end{figure}

However, if our idea  of electron-doped FeAs layers in the 1048-compound is correct, it should be possible to induce high $T_c$ values above 15~K also in the 1038-compound by electron doping instead of Pt-substitution in the FeAs-layers. Indeed we were able to synthesize (Ca$_{0.8}$La$_{0.2}$)(Fe$_{1-x}$Pt$_x$)$_{10}$(Pt$_3$As$_8$) with $x \approx$ 0.03 (La-1038), where the small Pt-substitution alone is not sufficient to induce superconductivity according to the phase diagram (Figure~\ref{fig:phasedia}). The 1038-structure with La-substitution at the Ca-sites was confirmed by Rietveld-refinement as shown in Figure~\ref{fig:rietveld}.

A special feature arises in the 1038-structure from the missing Pt-atom in the layer in contrast to the 1048-structure (see Figure~\ref{fig:structures}). Calcium atoms below (and above) this vacancy are eightfold coordinated by arsenic atoms, which is not possible in the 1048-structure, where the vacancy is filled by the additional platinum atom. This favorable coordination increases the lattice energy, which is probably the reason why the 1038-compounds form only one polymorph, while the 1048-compounds form at least three and are much more affected by stacking disorder. However, the crystal structure determination of La-1038 clearly shows that lanthanum has a distinct preference to this site, where the higher charge of La$^{3+}$ further increases the lattice energy.

\begin{figure}
\center{
\includegraphics[width=80mm]{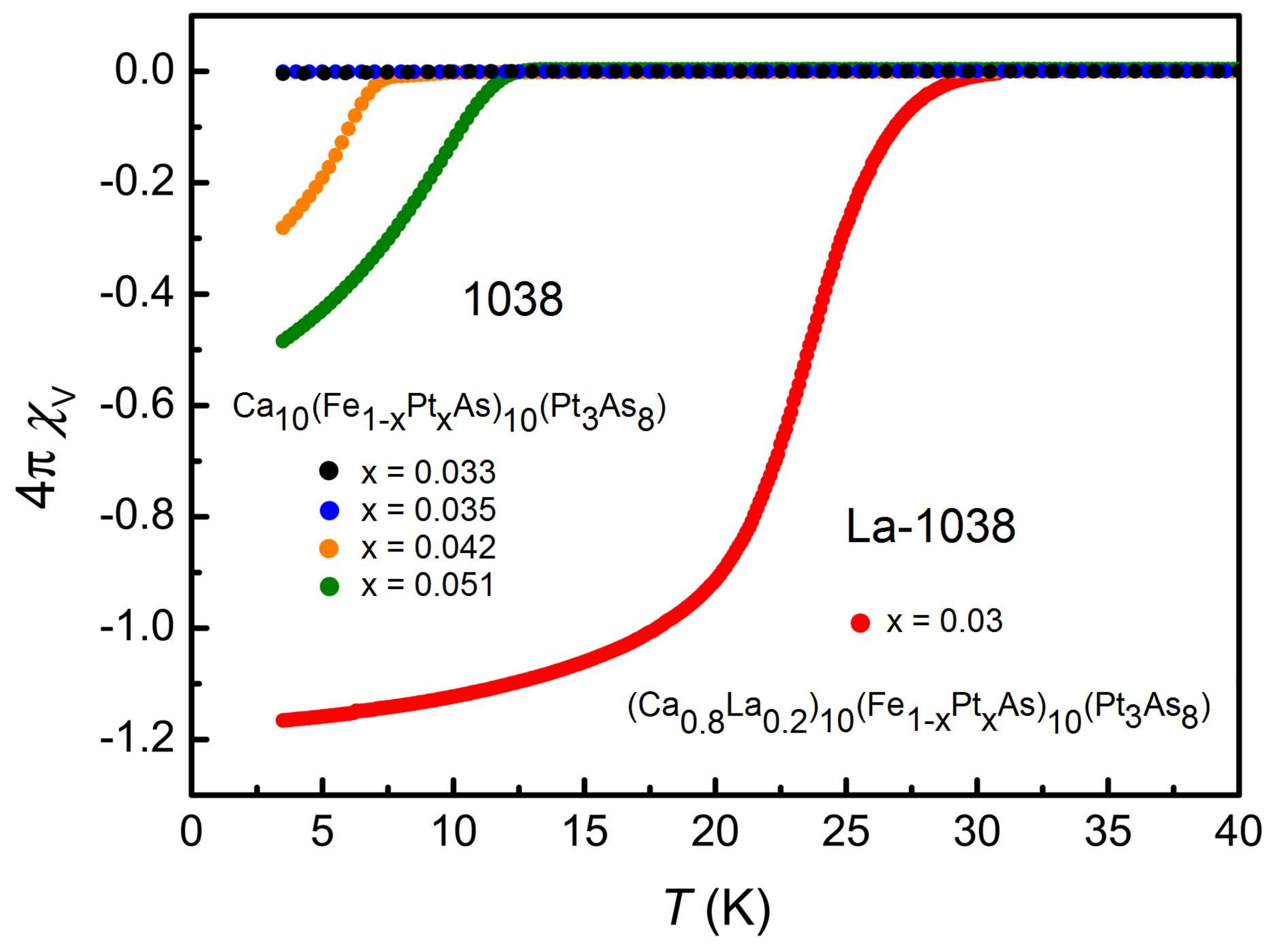}
\caption{AC-susceptibilty measurements of 1038-compounds with different Pt-substitution and of La-1038 with small Pt-substitution that is not sufficient to induce superconductivity.}
\label{fig:ACsus}
}
\end{figure}

Figure~\ref{fig:ACsus} shows AC-susceptibilty data of the 1038-compounds. No superconductivity emerges at low Pt-substitution ($x < $ 0.035), and $T_c$ remains below 15~K at $x$ = 0.051. In stark contrast to this, superconductivity is observed at 30~K in the La-1038 compound, where the Pt-substitution ($x$ = 0.03) is much too small to induce superconductivity.  Note also that the $T_c$ of La-1038 perfectly fits to the values of the 1048-compounds in the phase diagram (Figure~\ref{fig:phasedia}), which illustrates that only charge doping is crucial, regardless whether the structure is of the 1038-type or 1048-type. Given that Pt-substitution reduces $T_c$ in the 1048-compounds, we expect even higher values around 40~K for La-1038 without Pt at the iron site, but we were not yet able to prepare this. However, the finding of superconductivity at 30~K in La-1038 is a convincing proof of our idea that higher $T_c$ ($>$ 15~K) in the 1038/1048-materials can only emerge if the iron-arsenide layers are free of platinum and charge-doped.

In conclusion, our results emphasize the extraordinary role of the buffer layers in the platinum-iron-arsenide compounds which, for the first time, contain a second negatively charged layer beyond FeAs, which competes for the electrons provided by the Ca$^{2+}$ ions. In spite of this and amazingly enough, all results indicate that the electronic situation of the FeAs layers in the 1038 and 1048-compounds is almost identical to the simpler iron-arsenide superconductors, thus the Pt$_{3}$As$_8$ or Pt$_{4}$As$_8$ layers with its own special structures obviously attract just the proper amount of electrons to establish this situation. Given this as a parlor trick of nature, it is easy to accept that the charge balance between the layers may be delicate and can be manipulated in various ways. It is not surprising that the FeAs layer is much more susceptible to additional electrons than the Pt$_3$As$_8$ layer, because the states close to the Fermi-level are predominantly from iron arsenide. Thus extra electrons donated from La$^{3+}$ ions in La-1038 move to the FeAs layer and cause electron doping. Based on our results we can satisfactory explain the at first sight puzzling behavior of the critical temperatures in this new superconductors. These materials open new perspectives for future studies especially with respect to the detailed role and possible manipulations of the charge balance between the two negatively charged layers.

\begin{acknowledgments}

This work was financially supported by the German Research Foundation (DFG), Grant No. JO257/6-1.

\end{acknowledgments}

\bibliographystyle{apsrev}

%\bibliography{La1038}%

\begin{thebibliography}{14}
\expandafter\ifx\csname natexlab\endcsname\relax\def\natexlab#1{#1}\fi
\expandafter\ifx\csname bibnamefont\endcsname\relax
  \def\bibnamefont#1{#1}\fi
\expandafter\ifx\csname bibfnamefont\endcsname\relax
  \def\bibfnamefont#1{#1}\fi
\expandafter\ifx\csname citenamefont\endcsname\relax
  \def\citenamefont#1{#1}\fi
\expandafter\ifx\csname url\endcsname\relax
  \def\url#1{\texttt{#1}}\fi
\expandafter\ifx\csname urlprefix\endcsname\relax\def\urlprefix{URL }\fi
\providecommand{\bibinfo}[2]{#2}
\providecommand{\eprint}[2][]{\url{#2}}

\bibitem[{\citenamefont{L\"ohnert et~al.}(2011)\citenamefont{L\"ohnert,
  St\"urzer, Tegel, Frankovsky, Friederichs, and Johrendt}}]{Loehnert-2011}
\bibinfo{author}{\bibfnamefont{C.}~\bibnamefont{L\"ohnert}},
  \bibinfo{author}{\bibfnamefont{T.}~\bibnamefont{St\"urzer}},
  \bibinfo{author}{\bibfnamefont{M.}~\bibnamefont{Tegel}},
  \bibinfo{author}{\bibfnamefont{R.}~\bibnamefont{Frankovsky}},
  \bibinfo{author}{\bibfnamefont{G.}~\bibnamefont{Friederichs}},
  \bibnamefont{and} \bibinfo{author}{\bibfnamefont{D.}~\bibnamefont{Johrendt}},
  \bibinfo{journal}{Angew. Chem. Int. Ed.} \textbf{\bibinfo{volume}{50}},
  \bibinfo{pages}{9195} (\bibinfo{year}{2011}).

\bibitem[{\citenamefont{Ni et~al.}(2011)\citenamefont{Ni, Allred, Chan, and
  Cava}}]{Cava-PNAS2011}
\bibinfo{author}{\bibfnamefont{N.}~\bibnamefont{Ni}},
  \bibinfo{author}{\bibfnamefont{J.~M.} \bibnamefont{Allred}},
  \bibinfo{author}{\bibfnamefont{B.~C.} \bibnamefont{Chan}}, \bibnamefont{and}
  \bibinfo{author}{\bibfnamefont{R.~J.} \bibnamefont{Cava}},
  \bibinfo{journal}{Proc. Natl. Acad. Sci. U. S. A.}
  \textbf{\bibinfo{volume}{108}}, \bibinfo{pages}{E1019}
  (\bibinfo{year}{2011}).

\bibitem[{\citenamefont{Kakiya et~al.}(2011)\citenamefont{Kakiya, Kudo,
  Nishikubo, Oku, Nishibori, Sawa, Yamamoto, Nozaka, and Nohara}}]{Nohara-2011}
\bibinfo{author}{\bibfnamefont{S.}~\bibnamefont{Kakiya}},
  \bibinfo{author}{\bibfnamefont{K.}~\bibnamefont{Kudo}},
  \bibinfo{author}{\bibfnamefont{Y.}~\bibnamefont{Nishikubo}},
  \bibinfo{author}{\bibfnamefont{K.}~\bibnamefont{Oku}},
  \bibinfo{author}{\bibfnamefont{E.}~\bibnamefont{Nishibori}},
  \bibinfo{author}{\bibfnamefont{H.}~\bibnamefont{Sawa}},
  \bibinfo{author}{\bibfnamefont{T.}~\bibnamefont{Yamamoto}},
  \bibinfo{author}{\bibfnamefont{T.}~\bibnamefont{Nozaka}}, \bibnamefont{and}
  \bibinfo{author}{\bibfnamefont{M.}~\bibnamefont{Nohara}},
  \bibinfo{journal}{J. Phys. Soc. Jpn.} \textbf{\bibinfo{volume}{80}},
  \bibinfo{pages}{093704} (\bibinfo{year}{2011}).

\bibitem[{\citenamefont{Nishikubo et~al.}(2010)\citenamefont{Nishikubo, Kakiya,
  Danura, Kudo, and Nohara}}]{Nishikubo-2010}
\bibinfo{author}{\bibfnamefont{Y.}~\bibnamefont{Nishikubo}},
  \bibinfo{author}{\bibfnamefont{S.}~\bibnamefont{Kakiya}},
  \bibinfo{author}{\bibfnamefont{M.}~\bibnamefont{Danura}},
  \bibinfo{author}{\bibfnamefont{K.}~\bibnamefont{Kudo}}, \bibnamefont{and}
  \bibinfo{author}{\bibfnamefont{M.}~\bibnamefont{Nohara}},
  \bibinfo{journal}{J. Phys. Soc. Jpn.} \textbf{\bibinfo{volume}{79}},
  \bibinfo{pages}{095002} (\bibinfo{year}{2010}).

\bibitem[{\citenamefont{Nohara et~al.}(2012)\citenamefont{Nohara, Kakiya, Kudo,
  Oshiro, Araki, Kobayashi, Oku, Nishibori, and Sawa}}]{Nohara-2012}
\bibinfo{author}{\bibfnamefont{M.}~\bibnamefont{Nohara}},
  \bibinfo{author}{\bibfnamefont{S.}~\bibnamefont{Kakiya}},
  \bibinfo{author}{\bibfnamefont{K.}~\bibnamefont{Kudo}},
  \bibinfo{author}{\bibfnamefont{Y.}~\bibnamefont{Oshiro}},
  \bibinfo{author}{\bibfnamefont{S.}~\bibnamefont{Araki}},
  \bibinfo{author}{\bibfnamefont{T.~C.} \bibnamefont{Kobayashi}},
  \bibinfo{author}{\bibfnamefont{K.}~\bibnamefont{Oku}},
  \bibinfo{author}{\bibfnamefont{E.}~\bibnamefont{Nishibori}},
  \bibnamefont{and} \bibinfo{author}{\bibfnamefont{H.}~\bibnamefont{Sawa}},
  \bibinfo{journal}{Solid State Commun.} \textbf{\bibinfo{volume}{152}},
  \bibinfo{pages}{635} (\bibinfo{year}{2012}).

\bibitem[{\citenamefont{Neupane et~al.}(unpublished)\citenamefont{Neupane, Liu,
  Xu, Wang, Ni, Allred, Wray, Lin, Markiewicz, Bansil et~al.}}]{Neupane-2011}
\bibinfo{author}{\bibfnamefont{M.}~\bibnamefont{Neupane}},
  \bibinfo{author}{\bibfnamefont{C.}~\bibnamefont{Liu}},
  \bibinfo{author}{\bibfnamefont{S.-Y.} \bibnamefont{Xu}},
  \bibinfo{author}{\bibfnamefont{Y.~J.} \bibnamefont{Wang}},
  \bibinfo{author}{\bibfnamefont{N.}~\bibnamefont{Ni}},
  \bibinfo{author}{\bibfnamefont{J.~M.} \bibnamefont{Allred}},
  \bibinfo{author}{\bibfnamefont{L.}~\bibnamefont{Wray}},
  \bibinfo{author}{\bibfnamefont{H.}~\bibnamefont{Lin}},
  \bibinfo{author}{\bibfnamefont{R.}~\bibnamefont{Markiewicz}},
  \bibinfo{author}{\bibfnamefont{A.}~\bibnamefont{Bansil}},
  \bibnamefont{et~al.}, \bibinfo{journal}{arxiv:1110.4687}
  (\bibinfo{year}{unpublished}).

\bibitem[{\citenamefont{Coelho}(2007)}]{Topas}
\bibinfo{author}{\bibfnamefont{A.}~\bibnamefont{Coelho}},
  \emph{\bibinfo{title}{TOPAS-Academic, Version 4.1, Coelho Software}}
  (\bibinfo{address}{Brisbane}, \bibinfo{year}{2007}).

\bibitem[{\citenamefont{Blaha et~al.}(2001)\citenamefont{Blaha, Schwarz,
  Madsen, Kvasnicka, and Luitz}}]{Blaha-2001}
\bibinfo{author}{\bibfnamefont{P.}~\bibnamefont{Blaha}},
  \bibinfo{author}{\bibfnamefont{K.}~\bibnamefont{Schwarz}},
  \bibinfo{author}{\bibfnamefont{G.~K.~H.} \bibnamefont{Madsen}},
  \bibinfo{author}{\bibfnamefont{D.}~\bibnamefont{Kvasnicka}},
  \bibnamefont{and} \bibinfo{author}{\bibfnamefont{J.}~\bibnamefont{Luitz}},
  \emph{\bibinfo{title}{Wien2k - an augmented plane wave + local orbitals
  program for calculating crystal properties}} (\bibinfo{year}{2001}).

\bibitem[{\citenamefont{Schwarz and Blaha}(2003)}]{Schwarz-2003}
\bibinfo{author}{\bibfnamefont{K.}~\bibnamefont{Schwarz}} \bibnamefont{and}
  \bibinfo{author}{\bibfnamefont{P.}~\bibnamefont{Blaha}},
  \bibinfo{journal}{Comput. Mat. Sci.} \textbf{\bibinfo{volume}{28}},
  \bibinfo{pages}{259} (\bibinfo{year}{2003}).

\bibitem[{\citenamefont{Bader}(1990)}]{Bader-1990}
\bibinfo{author}{\bibfnamefont{R.~F.~W.} \bibnamefont{Bader}},
  \emph{\bibinfo{title}{Atoms in Molecules - A Quantum Theory}}
  (\bibinfo{publisher}{Oxford University Press}, \bibinfo{address}{London},
  \bibinfo{year}{1990}).

\bibitem[{\citenamefont{Cho et~al.}(unpublished)\citenamefont{Cho, Tanatar,
  Kim, Straszheim, Ni, Cava, and Prozorov}}]{Cho-2011}
\bibinfo{author}{\bibfnamefont{K.}~\bibnamefont{Cho}},
  \bibinfo{author}{\bibfnamefont{M.~A.} \bibnamefont{Tanatar}},
  \bibinfo{author}{\bibfnamefont{H.}~\bibnamefont{Kim}},
  \bibinfo{author}{\bibfnamefont{W.~E.} \bibnamefont{Straszheim}},
  \bibinfo{author}{\bibfnamefont{N.}~\bibnamefont{Ni}},
  \bibinfo{author}{\bibfnamefont{R.~J.} \bibnamefont{Cava}}, \bibnamefont{and}
  \bibinfo{author}{\bibfnamefont{R.}~\bibnamefont{Prozorov}},
  \bibinfo{journal}{arxiv:1111.1003}  (\bibinfo{year}{unpublished}).

\bibitem[{\citenamefont{Ren et~al.}(2008)\citenamefont{Ren, Lu, Yang, Yi, Shen,
  Li, Che, Dong, Sun, Zhou et~al.}}]{Ren-2008}
\bibinfo{author}{\bibfnamefont{Z.-A.} \bibnamefont{Ren}},
  \bibinfo{author}{\bibfnamefont{W.}~\bibnamefont{Lu}},
  \bibinfo{author}{\bibfnamefont{J.}~\bibnamefont{Yang}},
  \bibinfo{author}{\bibfnamefont{W.}~\bibnamefont{Yi}},
  \bibinfo{author}{\bibfnamefont{X.-L.} \bibnamefont{Shen}},
  \bibinfo{author}{\bibfnamefont{Z.-C.} \bibnamefont{Li}},
  \bibinfo{author}{\bibfnamefont{G.-C.} \bibnamefont{Che}},
  \bibinfo{author}{\bibfnamefont{X.-L.} \bibnamefont{Dong}},
  \bibinfo{author}{\bibfnamefont{L.-L.} \bibnamefont{Sun}},
  \bibinfo{author}{\bibfnamefont{F.}~\bibnamefont{Zhou}}, \bibnamefont{et~al.},
  \bibinfo{journal}{Chinese Physics Letters} \textbf{\bibinfo{volume}{25}},
  \bibinfo{pages}{2215} (\bibinfo{year}{2008}).

\bibitem[{\citenamefont{Rotter et~al.}(2008)\citenamefont{Rotter, Tegel, and
  Johrendt}}]{Rotter-2008}
\bibinfo{author}{\bibfnamefont{M.}~\bibnamefont{Rotter}},
  \bibinfo{author}{\bibfnamefont{M.}~\bibnamefont{Tegel}}, \bibnamefont{and}
  \bibinfo{author}{\bibfnamefont{D.}~\bibnamefont{Johrendt}},
  \bibinfo{journal}{Phys Rev Lett} \textbf{\bibinfo{volume}{101}},
  \bibinfo{pages}{107006} (\bibinfo{year}{2008}).

\bibitem[{\citenamefont{Sefat et~al.}(2008)\citenamefont{Sefat, Jin, McGuire,
  Sales, Singh, and Mandrus}}]{Sefat-2008}
\bibinfo{author}{\bibfnamefont{A.~S.} \bibnamefont{Sefat}},
  \bibinfo{author}{\bibfnamefont{R.}~\bibnamefont{Jin}},
  \bibinfo{author}{\bibfnamefont{M.~A.} \bibnamefont{McGuire}},
  \bibinfo{author}{\bibfnamefont{B.~C.} \bibnamefont{Sales}},
  \bibinfo{author}{\bibfnamefont{D.~J.} \bibnamefont{Singh}}, \bibnamefont{and}
  \bibinfo{author}{\bibfnamefont{D.}~\bibnamefont{Mandrus}},
  \bibinfo{journal}{Physical Review Letters} \textbf{\bibinfo{volume}{101}},
  \bibinfo{pages}{117004} (\bibinfo{year}{2008}).

\end{thebibliography}

\end{document}